\documentclass{article}
\usepackage[utf8]{inputenc}
\usepackage[T1]{fontenc}
\usepackage{amsmath,amsfonts,amssymb}
\usepackage{times}
\usepackage{xspace}
\usepackage{makeidx}
\usepackage{ifpdf}
\ifpdf
 \usepackage[colorlinks,pdftex]{hyperref}
\else
 \usepackage[ps2pdf,breaklinks=true,colorlinks=true,linkcolor=red,citecolor=green]{hyperref}
\fi

\newcommand{\Z}{\mathbb{Z}}

\makeindex

\begin{document}
\title{Sparse multivariate factorization by mean 
of a few bivariate factorizations.}
\author{Bernard.Parisse@univ-grenoble-alpes.fr}
\date{2016}


\maketitle

\bibliographystyle{abbrv}



\begin{abstract}
We describe an algorithm to factor sparse multivariate polynomials
using $O(d)$ bivariate factorizations where $d$ is the number of variables.
This algorithm is implemented in the Giac/Xcas computer algebra
system.
\end{abstract}

\section{Introduction}
To my knowledge, there are three classes of algorithms to factor multivariate
polynomials over $\Z$~:
\begin{itemize}
\item reduction to bivariate factorization and Hensel lifting 
(Von zur Gathen and Kaltofen \cite{von1985factoring},
Bernardin and Monagan \cite{bernardin1997efficient})
\item evaluation of one variable at a sufficiently large $z$, 
factorization and
reconstruction by writing coefficients in basis $z$ with symmetric
remainders (heuristic factorization)
\item Kronecker-like substitution (replace one of the variable
by another one to a sufficiently large power).
\end{itemize}
Bivariate factorization may be obtained by 
partial differential equation (Gao \cite{gao2003factoring}) 
or interpolation or one of the previous method.

We present here a method that is adapted to sparse factors, where the
previous method would require too many ressources. For example Hensel
lifting does not work if a leading coefficient of the factors
vanishes once evaluated to 0 at other variables. 
The usual trick to avoid this is to translate the origin, 
but this will densify the polynomial to be factored.

\section{The algorithm}
\subsection{Main idea}
Let $n\geq 2$ and $P(x,x_1,...,x_n)$ be a sparse square-free polynomial 
that we want to factor, assume that the factorization is~:
$$ P=P_1...P_k$$
The basic idea is replace all variables $x_1,...,x_n$ with
$t,t,...,t$ and factor the substituted bivariate polynomial $P_{t,...,t}$, 
then
compare with the factorization of $P_{t^2,t,...,t}$ 
(where $x_1,...,x_n$
are substituted by $t^2,t,t,...,t$ in $P$) or with $P_{t^3,t,...,t}$ or etc. 
If the factorization is sparse enough,
there is a good chance that the factors will be similar (same number
of monomials, same pattern in $x$, same value for the coefficients), and the
monomial power differences in $t$ will give us the $x_1$ contribution
to the monomials. Doing the same for $x_2$, ..., $x_n$ will give us
the reconstruction.

The details are a little more complicated, because we must take care of
the content of the substituted polynomials $P_{t^.,t,...,t}$ 
and of the order of the
monomials having the same $x$ powers in a given factor.
The next example that motivated the implementation in Giac/Xcas will
demonstrate the main idea, problems and solutions.

\subsection{Example}
The following example was discussed on the \verb|sage-devel| list, 
it was obtained with a random generation command returning
2 polynomials in 5 variables. We make the product and
try to factor it back. It was not factored
by Sage 7.4 (with Singular 4 inside), 
but was reported to be factored by magma in 3s.
\begin{verbatim}
A:=37324800000000*a^25*b^9*c^25*d^21*E^21 + 
186624000000000*a^20*b^9*c^25*d^24*E^21 + 
37324800000000*a^20*b^4*c^25*d^22*E^21 + 
12441600000000*a^20*b^4*c^28*d^21*E^18 + 
373248000000*a^16*b^4*c^25*d^21*E^20 + 
1866240000000*a^16*b^4*c^26*d^21*E^18 + 
186624000000*a^13*b^6*c^25*d^21*E^17 + 
12441600000*a^13*b^5*c^25*d^21*E^12 + 
3110400000*a^13*b^7*c^23*d^16*E^12 + 
12441600000*a^13*b^4*c^25*d^16*E^13 + 
3110400000*a^16*b^4*c^20*d^16*E^12 + 
622080000*a^13*b*c^21*d^16*E^12 + 
233280000*a^13*b*c^20*d^17*E^8 + 77760000*a^13*b*c^15*d^18*E^8 + 
25920000*a^13*b*c^15*d^14*E^10 + 25920000*a^13*b^4*c^15*d^10*E^8 + 
17280000*a^8*b*c^15*d^14*E^8 + 3240000*a^8*b^4*c^15*d^6*E^8 + 
216000*a^4*b^3*c^15*d^6*E^8 + 216000*a^4*b*c^10*d^9*E^8 + 
86400*a^4*b*c^10*d^8*E^7 + 32400*a^7*b*c^10*d^3*E^7 + 
2700*a^4*b^4*c^10*d^3*E^3 + 675*a^6*b*c^7*d^3*E^3 + 
1125*a^5*b*c^2*d^3*E^3 + 135*b^5*c^2*d^3*E^3 + 
27*c^2*d^6*E^3 + 12*c^7 + 9*c^3*d^3 + a^2;


B:=1105920000000000*E^36*a^7*b^16*c^6*d^33+
276480000000000*E^35*a^7*b^16*c^6*d^33+
20736000000000*E^32*a^2*b^15*c^6*d^33+
345600000000000*E^31*a^7*b^16*c^6*d^33+
69120000000000*E^29*a^7*b^15*c^8*d^33+
103680000000000*E^29*a^7*b^15*c^6*d^33+
10368000000000*E^26*a^2*b^15*c^6*d^33+7680000*E^21*a*b^9*d^18+
57600000000*E^20*a^7*b^14*c^2*d^29+
11520000000*E^20*a^6*b^12*c^2*d^29+
2400000000*E^20*a^6*b^9*d^29+
1728000000000*E^20*a^2*b^19*c^6*d^33+
216000000000*E^20*a^2*b^14*c^10*d^31+
864000000000*E^20*a^2*b^14*c^6*d^37+
216000000000*E^20*a^2*b^14*c^2*d^32+3840000000*E^20*a^2*b^14*d^29+
480000000*E^20*a*b^9*d^34+96000000*E^18*a*b^9*c^3*d^29+
76800000*E^18*a*b^9*d^31+28800000*E^17*a*b^9*c^5*d^26+
5760000*E^17*a*b^9*c^4*d^23+384000*E^17*b^14*d^16+
76800*E^17*b^12*d^16+1920000*E^17*b^9*d^22+
11520*E^14*b^4*d^13+46080*E^12*b^8*d^21+
38400*E^12*b^4*d^19+3840*E^8*b^6*d^13+768*E^8*b^4*c*d^11+
24*E^3*b^7*d^8+96*E^3*b^4*c^2*d^11+96*E^3*c^2*d^8+
6*E^3*d^8+2*b^2*d^3+3*b*d^7+c^5;
\end{verbatim}
The smallest partial degree of the product is 9+19 in $b$, therefore
$b$ will be our $x$ variable, while $a,c,d,E$ are 
our $x_1,...,x_4$ variables. $A$ and $B$ are
irreducible, we set $P=AB$.

Factoring $P(x,t,t,t,t)$ gives
\begin{verbatim}
t^5*
(37324800000000*b^9*t^90+186624000000000*b^9*t^88
+3110400000*b^7*t^62+186624000000*b^6*t^74+
12441600000*b^5*t^69+135*b^5*t^6+
37324800000000*b^4*t^86+12441600000000*b^4*t^85+
373248000000*b^4*t^80+1866240000000*b^4*t^79
+12441600000*b^4*t^65+3110400000*b^4*t^62+25920000*b^4*t^44
+3240000*b^4*t^35+2700*b^4*t^18+
216000*b^3*t^31+
622080000*b*t^60+233280000*b*t^56+77760000*b*t^52
+25920000*b*t^50+17280000*b*t^43+216000*b*t^29
+86400*b*t^27+32400*b*t^25+675*b*t^17+1125*b*t^11
+27*t^9+12*t^5+9*t^4+1)*...
\end{verbatim}
Factoring $P(x,t^2,t,t,t)$ gives
\begin{verbatim}
t^7*
(37324800000000*b^9*t^113+186624000000000*b^9*t^106
+3110400000*b^7*t^73+186624000000*b^6*t^85+
12441600000*b^5*t^80+135*b^5*t^4+
37324800000000*b^4*t^104+12441600000000*b^4*t^103+
373248000000*b^4*t^94+1866240000000*b^4*t^93+
15552000000*b^4*t^76+25920000*b^4*t^55+
3240000*b^4*t^41+2700*b^4*t^20+216000*b^3*t^33+
622080000*b*t^71+233280000*b*t^67+77760000*b*t^63+
25920000*b*t^61+17280000*b*t^49+216000*b*t^31
+32400*b*t^30+86400*b*t^29+675*b*t^21+1125*b*t^14+
27*t^7+12*t^3+9*t^2+1)*...
\end{verbatim}
It is clearly a similar factorization, the number of monomials differ only
by 1 ($12441600000*b^4*t^{65}+3110400000*b^4*t^{62}$ is grouped in one
monomial in the second factorization), and the order is not the same for 
the coefficient of $b$. Note that there is also a content term in $t$.
In fact, we just got an unlucky evaluation at $x_1=a=t^2$, 
$x_1=a=t^3$ is also unlucky, while $x_1=a=t^4$ returns 30 monomials like $t$.
\begin{verbatim}
t^9*
(37324800000000*b^9*t^161+186624000000000*b^9*t^144+
3110400000*b^7*t^97+186624000000*b^6*t^109+
12441600000*b^5*t^104+135*b^5*t^2+
37324800000000*b^4*t^142+12441600000000*b^4*t^141+
373248000000*b^4*t^124+1866240000000*b^4*t^123+
3110400000*b^4*t^106+12441600000*b^4*t^100+25920000*b^4*t^79+
3240000*b^4*t^55+2700*b^4*t^26+216000*b^3*t^39+
622080000*b*t^95+233280000*b*t^91+77760000*b*t^87+
25920000*b*t^85+17280000*b*t^63+32400*b*t^42+
216000*b*t^37+86400*b*t^35+675*b*t^31+1125*b*t^22+
27*t^5+t^2+12*t+9)*...
\end{verbatim}

In order to compare the two factorizations, we must solve these 3 problems:
content, number of monomials, and monomials ordering.

\subsection{Detailled method}
We assume that the factors of the bivariate factorization are $x$-distincts,
that is the distribution of non-zero coefficients in powers of $x$ are
not the same. This way, we can isolate the same factor in two bivariate
factorizations, and we will now reconstruct the true factor.

To solve the content normalization problem, 
we will as usual reconstruct the multiple of the factor of
$P$ that has the same leading coefficient (lc) as $P$ in $x$
(in the example the multiple of $A$ having same leading coefficient
as $P=AB$ in $b$).
It means that we can ignore the content 
in the factorization of $P_{t^.,...,t}$ and that we multiply 
the factor $f$ of $P_{t^.,...,t}$
by lc$(P)_{t^.,...,t}/$lc$(f)$. 
In our example the leading coefficient of $P$ is
\verb|193491763200000000000000000*a^22*c^31*d^54*E^41*(a^5+5*d^3)|
We can ignore the integer factor, therefore we multiply $P_{t,...,t}$ the
factor by
$t^{22+31+54+41}(t^5+5t^3)/(t^{90}+5t^{88})=t^{63}$. For $P_{t^4,...,t}$, we
multiply the factor
by $t^{4\times 22+31+54+41}(t^{20}+5t^3)/(t^{161}+5t^{144})=t^{73}$.

Solving the number of monomials is done like for any modular reconstruction:
if an evaluation with $x_k$ replaced by $t^j$ 
contains less term than the previous one, we ignore it 
(unlucky evaluation),
if an evaluation contains more terms than a previous one, we throw what
we had before and restart from this one. If the number of
monomials is the same, we also check
that the non-zero partials degrees in $x$ are the same.

Keeping the right order of monomials is more original: it can be 
done by comparing first the $x$
power, then comparing the coefficient of the monomial (it is
impossible to insure the same ordering by sorting with powers of $t$). 
In order to do that we must insure that in the factor to be reconstructed 
the coefficients of the monomials of the same power of $x$ are all
distincts. If this is not the case, we can dilate some
variables by a constant factor and retry (in our implementation
we dilate all variables except $x$ randomly by $\pm 1$ or $\pm 2$).

If we have two matching factors for evaluations at $t..t,t^j,t..t$ and
$t..t,t^{j'},t..t$ then the power of $x_k$ in a monomial is 
the difference of powers in $t$ of the same monomial in the
two factorizations divided by $j-j'$.
For example the first monomial in $P_{t,..,t}$ multiplied by $t^{63}$ is
$37324800000000*b^9*t^{90+63}$, the corresponding monomial in $P_{t^4,...t}$
multiplied by $t^{73}$ is $37324800000000*b^9*t^{161+73}$, that's a power
contribution for $x_1=a$ of $(234-153)/3=27$. Indeed the leading
coefficient of $A$ is $a^{25}$, multiplied
by $a^2$ inside the leading coefficient of $Q$ in $b$ is $a^{27}$.

\subsection{Implementation}
This algorithm is implemented in C++
in the file \verb|ezgcd.cc| of the source
code of Giac/Xcas, in the function 
\verb|try_sparse_factor_bi|
It factors the polynomial in the example in less than 2s (without 
this function, the factorization was impracticable).

We hope it will help other open-source softwares implement more efficient
sparse multivariate factorization algorithms!

\bibliography{gb}

\end{document}